\documentclass{article}
\usepackage{PRIMEarxiv} 

\usepackage[T1]{fontenc}      
\usepackage{lmodern}          
\usepackage{microtype}        

\usepackage{amsmath}          
\usepackage{amssymb}          

\usepackage{graphicx}         
\graphicspath{{media/}}       
\usepackage{array}            
\usepackage{booktabs}         
\usepackage{multirow}         
\usepackage{tabularx}         

\usepackage{algorithm}        
\usepackage{algorithmic}      

\usepackage{tikz}             
\usetikzlibrary{positioning, arrows.meta} 

\usepackage{ragged2e}         
\usepackage{enumitem}         
\usepackage{float}            
\usepackage{caption}          
\usepackage{nicefrac}         

\usepackage[hypertexnames=false]{hyperref} 

\usepackage{fancyhdr}
\title{\textbf{From Prompt--Response to Goal-Directed Systems:}\\
The Evolution of Agentic AI Software Architecture}
\author{
  Mamdouh Alenezi \\
  The Saudi Technology and Security Comprehensive Control Company "Tahakom" \\
  Riyadh \\
  Saudi Arabia\\
}

\begin{document}
\maketitle

\begin{abstract}
Agentic AI denotes an architectural transition from stateless, prompt-driven generative models toward goal-directed systems capable of autonomous perception, planning, action, and adaptation through iterative control loops. This paper examines this transition by connecting foundational intelligent agent theories, including reactive, deliberative, and Belief-Desire-Intention models, with contemporary LLM-centric approaches such as tool invocation, memory-augmented reasoning, and multi-agent coordination. The paper presents three primary contributions: (i) a reference architecture for production-grade LLM agents that separates cognitive reasoning from execution using typed tool interfaces; (ii) a taxonomy of multi-agent topologies, together with their associated failure modes and mitigation approaches; and (iii) an enterprise hardening checklist that incorporates governance, observability, and reproducibility considerations. Through an analysis of emerging industry platforms, including Kore.ai, Salesforce Agentforce, TrueFoundry, ZenML, and LangChain, the study identifies a convergence toward standardized agent loops, registries, and auditable control mechanisms. It is argued that the subsequent phase of agentic AI development will parallel the maturation of web services, relying on shared protocols, typed contracts, and layered governance structures to support scalable and composable autonomy. The persistent challenges related to verifiability, interoperability, and safe autonomy remain key areas for future research and practical deployment.
\end{abstract}

\begin{figure}[hbt!]
\centering
\includegraphics[width=0.98\textwidth]{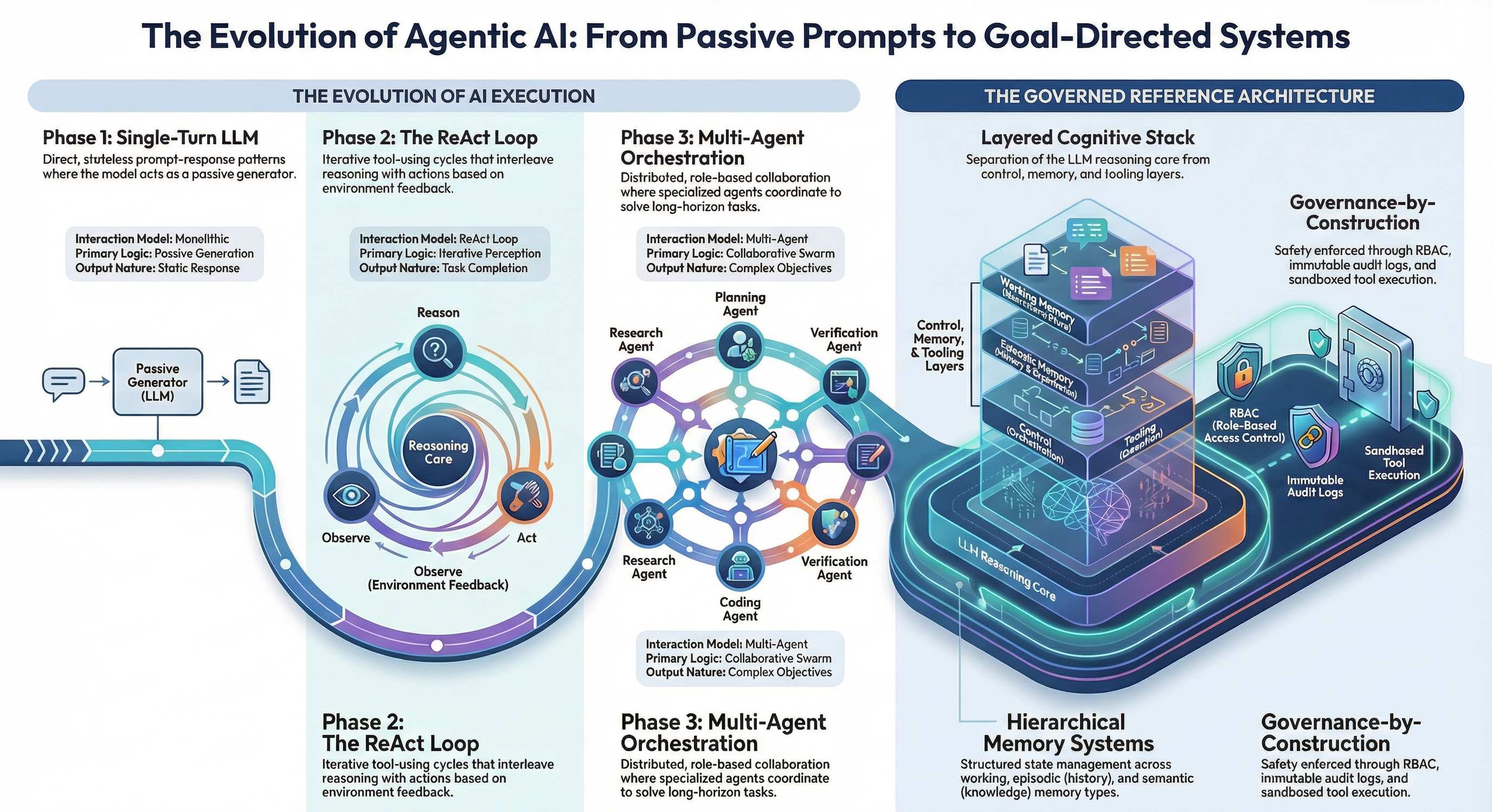}
\caption{Visual Abstract: From Passive Models to Goal-Directed Systems.}
\label{fig:va}
\end{figure}


\section{Introduction}

The earliest integrations of generative AI into software followed a largely stateless prompt-response pattern, with language models acting as passive text generators invoked inside fixed application boundaries \cite{abou2025agentic}. That pattern proved serviceable for content generation and simple Q\&A, but it is brittle for real-world operational workloads where tasks span multiple steps, external tool interfaces change over time, or regulatory requirements demand tamper-evident audit trails \cite{dwivedi2025agentic}. In practice engineers wrapped powerful models with fragile scaffolding—manual prompt chaining, external state managers, ad-hoc retry logic—to compensate for architectural gaps rather than address their root causes \cite{bandi2025rise}.

A consequential reconfiguration is now under way: agentic AI systems reframe the model as a cognitive kernel inside a closed-loop control architecture \cite{pati2025agentic, murugesan2025rise}. Unlike stateless invocation, an agentic architecture preserves persistent state across interactions, formulates and revises executable plans via typed tool interfaces, incorporates feedback from its environment to adapt behavior, and enforces governance constraints at runtime \cite{derouiche2025agentic}. Importantly, agency here is an architectural capability, not anthropomorphic intent; it arises from a clean separation of cognition from execution, state management, and policy enforcement.

This shift has both strategic and infrastructural dimensions. At the enterprise level the idea of an "Agentic Enterprise" reframes the AI component from a subordinate service to an orchestrating interface that mediates human intent and computational action, as seen in emerging product strategies such as Salesforce Agentforce \cite{salesforce_agentforce_2026}. At the platform level, production readiness is driving investments in observability, governed deployment, and reproducible pipelines—LangChain emphasizes traces as first-class observability artifacts \cite{langchain_blog_2026}, gateway-centric deployment models focus on policy-driven traffic control, and pipeline-first tooling targets reproducibility and auditability \cite{zenml_platform_2025}.

Architecturally, teams confront a pragmatic design choice along a spectrum. On one end, Symbolic or classical designs externalize planning into deterministic modules and explicit state machines, enabling verification at the cost of flexibility. On the other end, Neural or generative designs leverage stochastic LLM outputs for planning and decomposition, gaining adaptability but sacrificing predictability. Increasingly, production systems adopt hybrid patterns that use LLMs for high-level decomposition while enforcing symbolic constraints on tool execution so as to balance safety and adaptability \cite{murugesan2025rise, abou2025agentic}.

Beneath these choices lies a critical engineering substrate often overlooked in discussions that focus solely on model capability. Memory and serving architectures shape reliability and cost in production: Key-Value cache policies determine context retention under load, vector database latency constrains perception cycles, and context-window budgeting shapes both cost and effective reasoning. These systems-level constraints must be co-designed with cognitive components rather than treated as afterthoughts; platform investments in gateway architectures, memory tiering, and execution sandboxes reflect this co-design imperative.

This paper adopts an architectural lens to synthesize the technical, strategic, and operational dimensions of agentic AI. We center three research questions: what software primitives and design patterns define agentic architectures beyond prompt engineering; how do architectures evolve from single-agent control loops to coordinated multi-agent topologies; and which reliability, security, and governance mechanisms are essential for production deployment at scale. In response, we present a reference architecture that separates cognitive reasoning, hierarchical memory, typed tool invocation, and embedded governance; a taxonomy of multi-agent patterns with mapped failure modes and mitigations; and a practical enterprise hardening checklist linking observability, policy enforcement, and reproducibility to governance pillars spanning organization, compliance, operations, and security. Collectively, these contributions argue that the maturation of agentic AI will follow the trajectory of web services: not by model improvements alone, but through shared protocols, typed contracts, and layered governance that enable composable autonomy at scale.

\section{Related Work and Positioning}

The design of goal directed AI systems rests on decades of research into reactive, deliberative, and hybrid control architectures. Reactive architectures map perceptions directly to actions via condition–action rules or learned policies, offering low latency but brittleness when tasks require look ahead or reasoning about hidden state \cite{Stavrou2026GenAgenticAIReliability}. Deliberative architectures maintain explicit world models and use search/planning to choose actions, excelling at explainability and goal consistency but suffering from latency and model mismatch failures \cite{GonzalezSantamarta2025CognitiveArchitectures}. Hybrid architectures combine the two, using reactive layers for tight control loops and deliberative layers for goal setting and re planning; their reliability hinges on well defined interfaces, escalation rules, and time budgets \cite{Grosvenor2025HybridIntelligence}. The Belief–Desire–Intention (BDI) model provides a principled framework for structuring agency: beliefs correspond to world state and memory, desires to goals and constraints, and intentions to adopted plans and tool calls \cite{Frering2025BDILLM}. BDI’s commitment strategies and intention revision policies offer a clear control skeleton that modern generative agents can inherit, separating free form generation from governed behavior. These classical foundations continue to inform the architecture of today’s LLM based agents, ensuring that reliability is built into the control flow rather than bolted on afterwards.

Modern LLM agents extend the classical loop by integrating generative models as reasoning substrates. Tool use is a core capability, enabling agents to retrieve grounding information, perform actions, and interact with external environments. Recent LLMs natively support function calling, deciding whether to invoke a tool, selecting the appropriate one, and generating required parameters \cite{Mohammadi2025LLMAgents}. Planning and reasoning are essential for multi step tasks; the ReAct paradigm exemplifies this by interleaving reasoning steps with tool usage, allowing agents to adapt their actions in response to evolving context \cite{Xu2025LLMAgentsToolLearning}. Subsequent work has refined this pattern with explicit planning then execution cycles, self reflection (e.g., Reflexion), and tree based search (e.g., Tree of Thoughts) \cite{Ren2025SessionAwareAgents}. Evaluation of these capabilities has evolved beyond static benchmarks toward dynamic, execution based metrics that assess tool selection accuracy, parameter correctness, and progress rates in real time trajectories. Together, these advances transform LLMs from stateless text generators into agents that can decompose complex problems, sequence tool calls, and correct their own mistakes through iterative reasoning.

Scaling from single agents to teams introduces coordination challenges that are addressed by a variety of topological patterns. Surveys of LLM based multi agent systems categorize collaboration structures as peer to peer, centralized (orchestrator worker), or distributed, and coordination strategies as role based, rule based, or model based \cite{Tran2025MultiAgentLLMs}. In practice, orchestrator worker hierarchies assign a supervisor agent that decomposes tasks and delegates subtasks to specialized workers; debate/critique protocols allow agents to argue over proposals before reaching consensus; hierarchical setups nest agents with different authority levels; and swarm/market models use competitive or cooperative bidding to allocate resources. These mechanisms enable groups of LLM agents to share knowledge, parallelize subtasks, and align efforts toward shared objectives, effectively pooling expertise and mitigating individual model limitations \cite{Sami2026HumanAICollab}. However, coordination failures—such as miscommunication, deadlock, or collusion—remain active research areas, necessitating rigorous evaluation of multi agent collaboration in dynamic, open ended environments.

Deploying LLM agents in production demands rigorous engineering practices that cover observability, evaluation, security, and reproducibility. Observability tools provide tracing of LLM request response cycles, prompt version tracking, and agent/tool monitoring, enabling developers to diagnose failures and optimize performance. Evaluation frameworks systematically assess agent capabilities (planning, tool use, memory, collaboration) as well as reliability, safety, and alignment, using both synthetic benchmarks and real world datasets. Security threats, notably prompt injection attacks, manipulate LLMs to operate outside safe boundaries; surveys classify these attacks by trust boundary violation and recommend socio technical defenses \cite{Geng2026WhiteBoxPrompt}. Policy enforcement requires runtime monitors that check for compliance, fairness, and safety invariants, while reproducibility is supported by workflow orchestration platforms (e.g., ZenML, LangChain) that version prompts, tools, and execution contexts \cite{Pahune2025LLMOps}. Together, these practices form an LLMOps stack that transforms experimental agents into governed, auditable, and scalable production systems.

The evolution from prompt response loops to goal directed agentic systems represents a convergence of classical agent theory with modern LLM centric practice. Architectural priorities are increasingly standardizing around typed tool interfaces, explicit control loops, and multi agent coordination protocols. Nevertheless, open problems remain in verifiability (how to formally certify agent behavior), interoperability (how to compose agents across different platforms), and safe autonomy (how to ensure agents remain aligned under open ended deployment). The next phase of agentic architecture is likely to resemble the maturation of web services: shared protocols, typed contracts, and layered governance that enable composable autonomy at scale. Addressing these challenges will require continued collaboration between the AI research community and industry practitioners, leveraging both classical principles and novel LLM driven innovations.

\section{From Reactive Models to Agentic Systems and a Reference Architecture}

The approach began with a targeted review of classical intelligent agent literature to ground the discussion in well established paradigms such as reactive control, deliberative planning, and Belief-Desire-Intention models. Foundational works provided the vocabulary and analytical levers we reuse when mapping older agent concepts to modern LLM-centric patterns \cite{bratman_1987,rao_georgeff_1995,wooldridge_2002,brooks_1986,russell_norvig_2020}. Complementing that historical layer, we surveyed contemporary research on tool use, retrieval-augmented generation, and agent loops to capture the mechanisms by which language models acquire external capabilities, persistent context, and iterative decision-making behaviors \cite{lewis_rag_2020,yao_react_2023,schick_toolformer_2023,shinn_reflexion_2023}.

To bridge academic and operational perspectives, we systematically collected vendor and platform materials that describe agent orchestration, tool interfaces, deployment models, and observability features. These sources include publicly available product briefs, technical blog posts, API documentation, and vendor positioning statements from a set of representative providers \cite{koreai_platform_2026,salesforce_agentforce_2026,truefoundry_agentic_2025,zenml_platform_2025,langchain_blog_2026,bain_home_2026}. We treat those items as grey literature: they are informative for identifying real-world constraints, common design patterns, and engineering priorities, but they are not used as evidence for comparative performance claims. Where platform descriptions made concrete architectural assertions, we extracted the claims verbatim and annotated them in the text so readers can distinguish descriptive reporting from analytic inference.

The conceptual foundation of agentic AI predates LLMs by decades. Classic AI treated an \emph{agent} as an entity that perceives an environment and acts to achieve goals \cite{russell_norvig_2020}. Architectures historically spanned reactive control, deliberative planning, and hybrid combinations. Reactive control denotes tightly coupled perception-action loops with minimal internal state, exemplified by subsumption-style layered control in robotics \cite{brooks_1986}. Deliberative planning relies on explicit internal world models and symbolic planning, often computationally expensive but interpretable. Hybrid systems combine reactive layers that provide fast safety and control with deliberative layers that perform planning, balancing robustness and flexibility. BDI (Belief-Desire-Intention) architectures formalized an agent's reasoning in terms of informational state (beliefs), objectives (desires), and committed plans (intentions) \cite{bratman_1987,rao_georgeff_1995}. Multi-agent systems research explored coordination, negotiation, and emergent behavior among interacting agents \cite{wooldridge_2002}. Together, these paradigms established the sense-think-act loop and the notion that intelligence is partly a property of architecture - state, control flow, and interfaces - not only of a learned model.

Large language models introduced a practical mechanism for broad-domain natural language understanding and generation, but early integrations were often monolithic: one prompt in, one answer out. Once LLMs were asked to do things such as retrieve fresh data, update records, or run analyses, system architectures began to resemble compound AI systems in which a reasoning component orchestrates calls to tools and retrievers \cite{lewis_rag_2020,yao_react_2023,schick_toolformer_2023}. The key shift is that the model is no longer the entire application; it is a cognitive kernel embedded in a control loop that includes tool invocation such as search, code execution and APIs, external memory and retrieval, iterative planning and self-correction, and monitoring and governance. LangChain's recent writing emphasizes the rise of agent engineering as a discipline, focusing on behavior analysis at scale and the centrality of traces to understand agent decisions \cite{langchain_blog_2026}. This is consistent with a broader architectural lesson: when control flow is partially learned and stochastic, runtime observability becomes as important as static code inspection.

\begin{figure}[hbt!]
\centering
\includegraphics[width=0.8\textwidth]{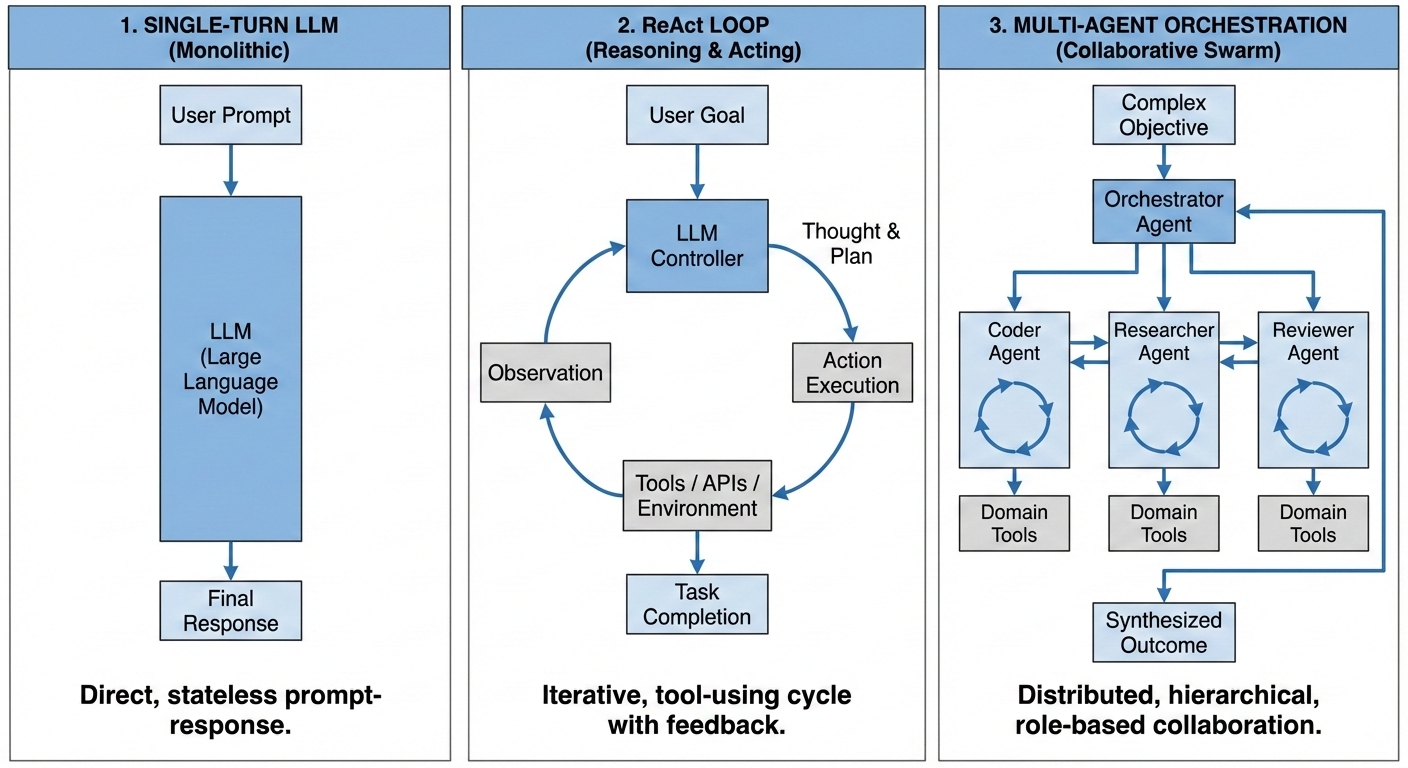}
\caption{Comparing the Single-Turn LLM vs. The ReAct Loop vs. Multi-Agent Orchestration.}
\label{fig:comparision}
\end{figure}

\begin{figure}[t]
\centering
\begin{tikzpicture}[
  node distance=6mm and 4mm,
  box/.style={draw, rounded corners, align=center, inner sep=4pt, minimum height=8mm, font=\small},
  layer/.style={draw, rounded corners, align=left, inner sep=5pt, minimum width=65mm, font=\small},
  arrow/.style={-{Latex[length=2.5mm]}, thick},
  every node/.append style={font=\small}
]
\node[box] (user) {Human\\(intent, constraints)};
\node[box, below=of user] (iface) {Agent Interface\\(chat, UI, API)};
\node[box, below=of iface] (core) {Agent Core\\LLM reasoning};

\node[layer, right=12mm of core] (control) {\textbf{Control Layer}\\
Planner/policy\\
State machine\\
Retry \& backoff\\
Circuit breakers};

\node[layer, below=5mm of control] (memory) {\textbf{Memory Layer}\\
Working context\\
Episodic store\\
Semantic KB/vector store\\
Preferences \& profiles};

\node[layer, below=5mm of memory] (tools) {\textbf{Tooling Layer}\\
Tool registry + schemas\\
Connectors/adapters\\
Sandboxed execution\\
RAG retrieval};

\node[layer, below=5mm of tools] (gov) {\textbf{Governance \& Observability}\\
RBAC, audit logs\\
Tracing \& evaluation\\
Policy enforcement\\
Cost \& rate limits};

\node[box, right=12mm of tools] (env) {External Environment\\(apps, data, web, infra)};

\draw[arrow] (user) -- (iface);
\draw[arrow] (iface) -- (core);
\draw[arrow] (core) -- (control);
\draw[arrow] (control) -- (memory);
\draw[arrow] (memory) -- (tools);
\draw[arrow] (tools) -- (gov);

\draw[arrow] (tools.east) -- (env.west);
\draw[arrow, bend left=30] (env.north) to (control.east);

\draw[arrow, dashed, shorten <=2pt] (gov.west) -- ++(-8mm,0) |- (core.west);
\draw[arrow, dashed, shorten <=2pt] (gov.west) -- ++(-8mm,0) |- (tools.west);
\end{tikzpicture}
\caption{Reference architecture for agentic AI systems. Cognition (LLM) is separated from control flow, memory, and tool execution; governance and observability cross-cut the stack. This separation reflects enterprise platform emphasis on orchestration, security, and tracing \cite{koreai_platform_2026,truefoundry_agentic_2025,zenml_platform_2025,langchain_blog_2026}.}
\label{fig:refarch}
\end{figure}
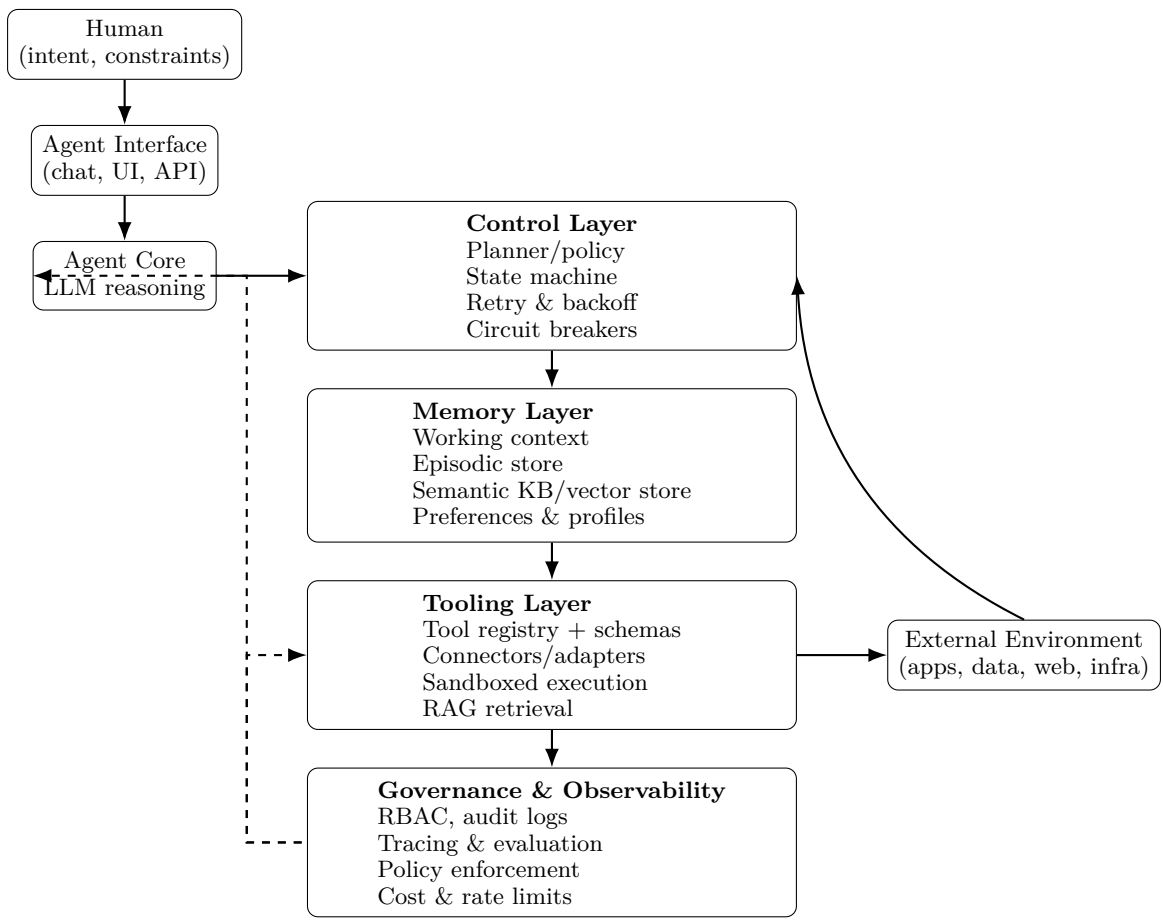

Figure~\ref{fig:refarch} presents a layered reference architecture for modern LLM-based agents, designed to separate concerns and to support governance-by-construction. At the top of the stack sits the human actor providing intent and constraints and the agent interface which can be chat, UI, or API. The Agent Core contains the LLM reasoning component. Separating cognition from execution, a control layer implements planner and policy logic, state machines, retry and backoff logic, and circuit breakers. A memory layer holds working context, an episodic store, semantic knowledge bases and vector stores, and user preferences and profiles. The tooling layer implements the tool registry and schemas, connectors and adapters, sandboxed execution environments, and retrieval-augmented generation. Governance and observability are cross-cutting and include RBAC and audit logs, tracing and evaluation, policy enforcement, and cost and rate limits. External environment integrations connect to applications, data, web resources, and infrastructure. Cognition (LLM) is intentionally separated from control flow, memory, and tool execution; governance and observability cross-cut the stack. This separation reflects enterprise platform emphasis on orchestration, security, and tracing \cite{koreai_platform_2026,truefoundry_agentic_2025,zenml_platform_2025,langchain_blog_2026}.

\begin{figure}[hbt!]
\centering
\includegraphics[width=0.8\textwidth]{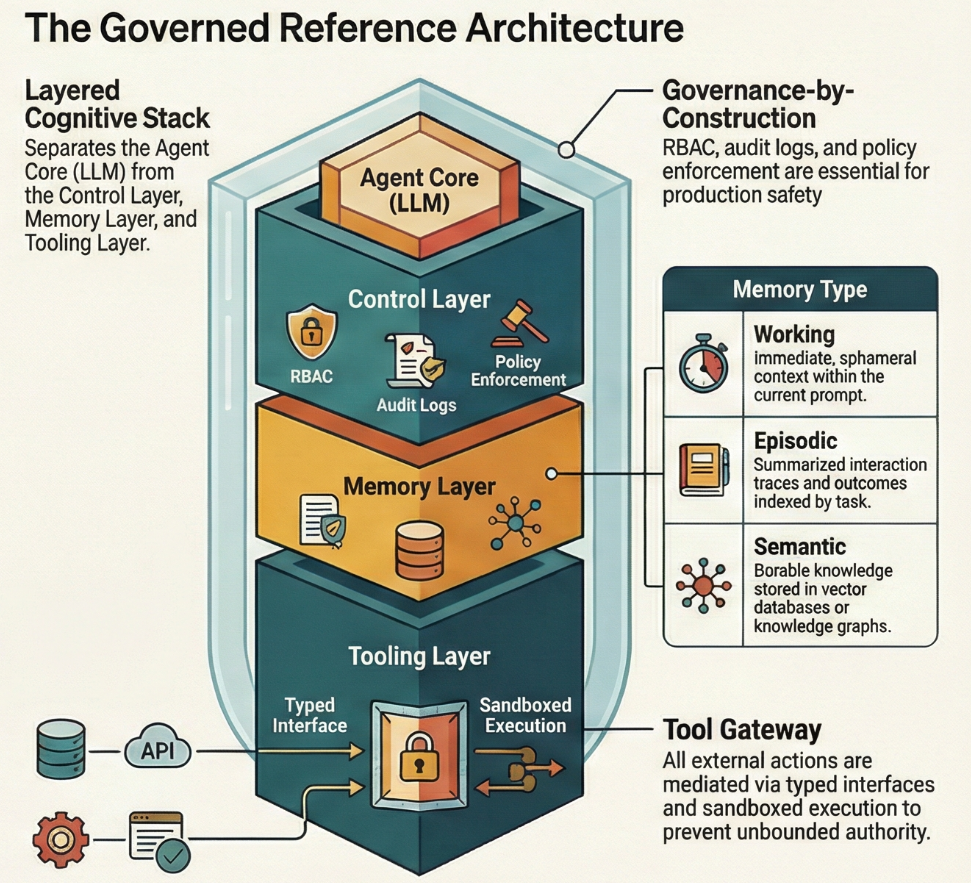}
\caption{The Governed Reference Architecture.}
\label{fig:ref}
\end{figure}

\textbf{Agent loop as a first-class control structure.} Agentic behavior arises from an iterative control loop rather than a single inference. Algorithm~\ref{alg:agentloop} sketches a generic loop aligned with ReAct-style interleaving of reasoning and acting \cite{yao_react_2023}.

\begin{algorithm}[h]
\caption{Generic agent loop (goal-directed tool-using agent)}
\label{alg:agentloop}
\begin{algorithmic}[1]
\REQUIRE user goal \(g\), tools \(T\) with schemas, memory \(M\), policies \(P\)
\STATE \(s \gets \text{InitState}(g)\)
\FOR{\(k = 1\) \TO \(K_{\max}\)}
  \STATE \(c \gets \text{BuildContext}(s, M, P)\)
  \STATE \(p \gets \text{PlanStep}(\text{LLM}, c)\) \COMMENT{propose next action or subgoal}
  \IF{ViolatesPolicy(\(p\), \(P\))}
    \STATE \(p \gets \text{RepairPlan}(p, P)\)
  \ENDIF
  \IF{IsToolCall(\(p\))}
    \STATE \(r \gets \text{ExecuteTool}(p, T)\) \COMMENT{typed invocation + sandbox}
    \STATE \(s \gets \text{UpdateState}(s, p, r)\)
    \STATE WriteMemory(\(M\), \(s\), \(p\), \(r\))
  \ELSE
    \RETURN FinalizeAnswer(\(p\), \(s\))
  \ENDIF
  \IF{ShouldStop(\(s\))}
    \RETURN SummarizeProgress(\(s\))
  \ENDIF
\ENDFOR
\RETURN FailSafe(\(s\)) \COMMENT{graceful degradation + escalation}
\end{algorithmic}
\end{algorithm}

Architecturally, the key is that tool execution is not part of the model; it is mediated by typed interfaces, sandboxes, and policies. This separation supports governance and auditability emphasized in enterprise agent platforms \cite{koreai_platform_2026,truefoundry_agentic_2025,salesforce_agentforce_2026}.

\textbf{Tool interfaces: typed, discoverable, and governable.} Tool use is the bridge between language and action. Early systems often used unstructured prompt patterns which proved fragile. Modern designs treat tools as typed contracts exposing schemas for inputs, outputs, and preconditions; as discoverable registry entries that are enumerated, versioned, and access-controlled; and as sandboxed execution units that run under least privilege with rate limits and safe defaults. TrueFoundry explicitly describes registries and schema validation and policy-oriented deployment of Model Context Protocol servers to manage agent traffic and enforce limits \cite{truefoundry_agentic_2025}. Kore.ai highlights tool and dialog integrations and admin controls plus governance capabilities such as RBAC, audit logs, and guardrails \cite{koreai_platform_2026}. These descriptions align with the trajectory toward tool registries as the agentic analogue of API gateways in microservices architectures.

\textbf{Memory: from context windows to hierarchical state.} Long-horizon behavior requires state beyond a single context window. Architecturally one can distinguish working memory as the immediate and ephemeral prompt context; episodic memory as summarized interaction traces and outcomes indexed by time and task; semantic memory as durable knowledge in documents, embeddings, or knowledge graphs; and preference or profile memory capturing user-specific constraints and style preferences when consent is present. Memory increases capability but introduces privacy and governance concerns, motivating policy-aware retrieval and access control. Enterprise platforms foreground governance primitives including RBAC, audit logs, and policy enforcement as integral architectural elements rather than afterthoughts \cite{koreai_platform_2026,truefoundry_agentic_2025,salesforce_agentforce_2026}. ZenML frames governance in workflow terms, emphasizing central key management, role-based access control, and audit-ready lineage from raw data to final outputs \cite{zenml_platform_2025}.

\textbf{Planning and self-correction pipelines.} Planning in agentic systems spans a spectrum from implicit planning where the plan simply emerges in the LLM output, to explicit planning in which a planner module produces a task graph and an executor runs it, to iterative reflection where the agent critiques intermediate results, repairs plans, and retries. ReAct provides an influential template: interleave reasoning with actions so that subsequent steps are grounded in observations \cite{yao_react_2023}. Reflection-based methods aim to reduce error accumulation by revising plans based on outcomes \cite{shinn_reflexion_2023}. From an architectural viewpoint the important point is that planning is a pipeline not a prompt; the pipeline can be instrumented, evaluated, and constrained. LangChain's emphasis on deep agents and context management reflects this shift: as task length grows systems must manage context and provide debugging via structured traces and analysis tools \cite{langchain_blog_2026}.

\textbf{Observability: traces as the new debugging substrate.} Agent behavior is stochastic and tool-dependent which makes traditional unit testing insufficient. Observability therefore becomes a primary architecture requirement. Systems must capture detailed traces of each step including prompts, tool calls, responses, latencies, and errors; they must run regression suites on representative tasks and measure tool-use correctness; and they must monitor cost drivers such as tokens, tool API charges, and GPU utilization for self-hosted models. TrueFoundry highlights full agent observability covering prompt-to-tool execution and OpenTelemetry-compatible integration with Grafana, Datadog, and Prometheus \cite{truefoundry_agentic_2025}. Kore.ai describes built-in control and observability via tracing, analytics, audit logs, and governance \cite{koreai_platform_2026}. LangChain materials emphasize debugging and understanding agent behavior at scale via traces \cite{langchain_blog_2026}. These themes converge on a common architectural claim: without traces, agent systems cannot be engineered reliably.

Transforming a reference architecture into an executable specification requires formalizing architectural intent as enforceable invariants that constrain system behavior across implementations. The proposed separation of concerns between LLM cognition, orchestration and control, tools, and memory becomes operational only when encoded as normative requirements expressed through mandatory and recommended clauses. For instance, all side-effecting actions must be mediated by a policy enforcement gateway, thereby preventing direct tool invocation by the language model and ensuring that authorization, compliance, and risk evaluation precede any external interaction. Similarly, every tool invocation should be strongly typed and versioned, with schemas and tool versions recorded as part of execution metadata. This requirement establishes reproducibility and auditability, enabling deterministic replay and longitudinal evaluation across model and tool upgrades. By codifying these constraints, the architecture evolves from a conceptual layering model into a verifiable contract that governs how intelligent agents interact with external systems.

Operational accountability further depends on standardized observability and bounded autonomy. Each execution run must produce a trace that captures a minimal yet sufficient set of provenance attributes, including model identifier, prompt version, tool versions, policy decisions, memory operations, principal identity, and resource budgets. Such traces support governance, incident response, and empirical research by providing a consistent substrate for monitoring and evaluation. In parallel, budgeted autonomy should be treated as a first-class invariant: systems must enforce explicit limits on tokens, execution time, tool invocations, and monetary cost, and must define fail-safe termination behavior when limits are exceeded. This requirement formalizes safe degradation and prevents unbounded agentic behavior. Collectively, these invariants elevate the architecture into a normative specification that can be implemented, audited, and benchmarked across organizations, thereby enabling interoperability, trust, and regulatory alignment in the deployment of AI agent systems.

\section{Production Hardening: Governance, Reliability, and Security}
The transition of agentic AI from research prototypes to enterprise production systems necessitates a fundamental architectural shift towards intrinsic safety and operational robustness. Unlike deterministic software, agentic systems possess a dynamic, generative core that expands the potential blast radius of failures—a single hallucination can cascade into an incorrect database write, and a prompt injection can escalate into a privileged action. Consequently, production hardening is not merely an add-on but must be designed as foundational, interlocking architectural layers encompassing governance, reliability, and security. Enterprise platforms reflect this imperative by embedding these concerns as first-class primitives, signaling that agentic systems must integrate into existing organizational risk and compliance frameworks, not exist as experimental outliers.

Central to this architecture is a security and access control model built on least-privilege principles. This requires a robust identity layer that binds every action to a verifiable principal—whether a human user, a service account, or a specific agent role—followed by strict authorization enforcement using Role-Based (RBAC) or Attribute-Based Access Control (ABAC) for tools and data stores. A critical technical challenge is secret management, preventing the leakage of API keys and credentials into the model's context. Platforms like Kore.ai explicitly architect for this, listing AI security, guardrails, RBAC, and comprehensive audit logs as core features \cite{koreai_platform_2026}. Similarly, TrueFoundry emphasizes single sign-on (SSO), RBAC, and immutable audit logging as non-negotiable governance primitives \cite{truefoundry_agentic_2025}. Architecturally, this collective emphasis implies that agent stacks require a dedicated authorization plane, analogous to the zero-trust security models adopted in modern microservices ecosystems, through which all agent-tool and inter-agent communication must be mediated.

Agentic systems introduce novel reliability challenges distinct from traditional deterministic services. LLM-powered agents can enter unbounded loops, drift from their objective, or persistently select suboptimal actions. Mitigating these failures requires patterns adapted from distributed systems engineering but tailored for non-deterministic actors. These include imposing bounded loops with maximum step limits \(K_{\max}\) and tool-call caps; implementing circuit breakers to halt execution when error rates spike or unsafe action sequences are detected; designing tools to be idempotent to prevent duplicate side effects from retries; and integrating mandatory human-in-the-loop approval gates for high-risk actions such as financial transactions or data deletions. These controls are consistently emphasized across industry platforms as part of admin controls and policy enforcement layers \cite{koreai_platform_2026,truefoundry_agentic_2025}. Their adoption reflects a broader enterprise reality: agentic capabilities must be seamlessly integrated into established operational risk frameworks, a theme echoed in enterprise consulting guidance that prioritizes governance and sovereignty alongside capability \cite{bain_home_2026}.

When agentic systems process regulated data, compliance, auditability, and sovereignty translate directly into architectural mandates. This necessitates immutable, granular audit logs that record the complete chain of causality: which principal initiated an action, through which agent, using which model and prompt version, at what time, and with what outcome. Data residency requirements demand architectures that ensure data never leaves designated environments, such as Virtual Private Clouds (VPC), on-premises infrastructure, or specific geopolitical regions. Model governance requires rigorous versioning, approval workflows, and evaluation prior to deployment. TrueFoundry's architecture highlights deployment in VPC, on-prem, or air-gapped environments, stressing that "no data leaves your domain" while supporting compliance with standards like SOC 2, HIPAA, and GDPR \cite{truefoundry_agentic_2025}. ZenML similarly emphasizes enterprise deployment within the customer's environment and a strong compliance posture (SOC2, ISO 27001), framing the platform as a control plane where compute and data remain sovereign \cite{zenml_platform_2025}. This architectural focus on sovereignty aligns with strategic leadership concerns about regulatory and geopolitical constraints \cite{bain_home_2026}.

Finally, the non-deterministic nature of LLMs makes reproducibility and systematic change management paramount. Agent behavior is a complex function of prompts, tool definitions, retrieval corpora, and model versions. Ensuring reproducibility requires: version-controlled prompts and tools with tracked changes and rollback capabilities; comprehensive artifact lineage that records datasets, retrieval snapshots, and intermediate results; and captured execution environments (e.g., containers) to enable deterministic re-runs. ZenML addresses this by emphasizing artifact and environment versioning to solve "it worked on my machine" problems and facilitate safe rollbacks \cite{zenml_platform_2025}. TrueFoundry describes centralized registries for prompt lifecycle management \cite{truefoundry_agentic_2025}, while Kore.ai's platform includes dedicated tooling (Prompt Studio, Evaluation Studio, Model Hub) to support iterative, governed improvement \cite{koreai_platform_2026}. These elements point toward a converging enterprise architecture where agentic AI inherits and extends DevOps discipline—Continuous Integration, Continuous Deployment (CI/CD), and observability—to a new domain of versioned prompts, tool graphs, and retrieval contexts. The ultimate architectural implication is clear: production-ready agentic systems are not just advanced AI models but are complex, governed software platforms where reliability, security, and auditability are woven into the very fabric of their design.

\begin{figure}[hbt!]
\centering
\includegraphics[width=0.94\textwidth]{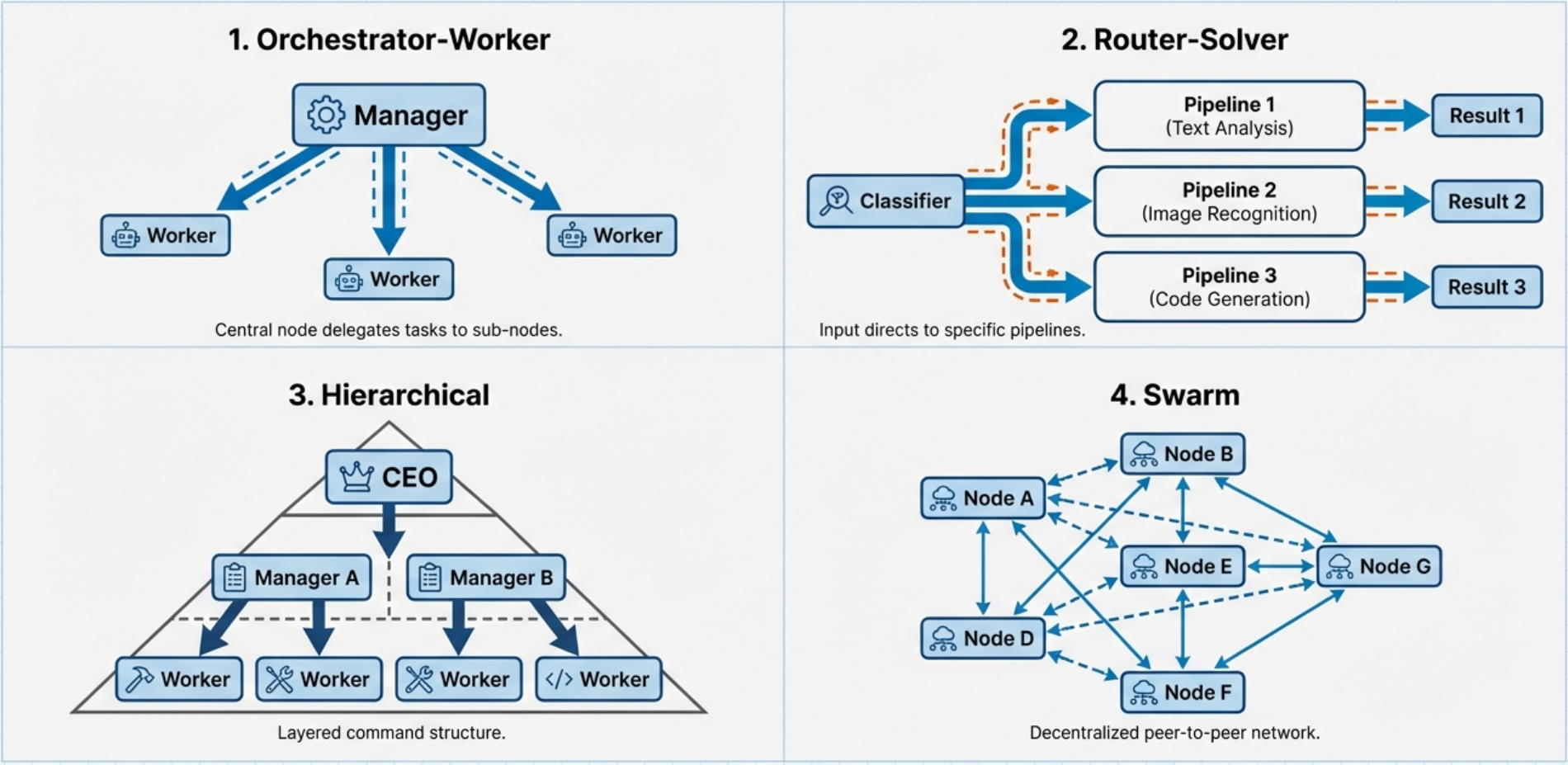}
\caption{Taxonomy of multi-agent topologies.}
\label{fig:taxonomy}
\end{figure}

\section{Multi-Agent Architectures: From Monoliths to Orchestrated Teams}
The evolution of agentic AI has revealed inherent limitations in single-agent paradigms, prompting a shift towards more distributed and collaborative computational models. While capable of handling well-defined tasks, monolithic single agents often struggle with context pollution—the progressive dilution and contamination of a fixed context window with extraneous information across long interactions. They further face tool overload, where a single agent must master an ever-expanding and potentially contradictory arsenal of functions, and they frequently falter at reliable long-horizon reasoning and planning. In response, multi-agent architectures have emerged as a foundational advance, distributing cognitive labor across specialized entities and introducing explicit coordination mechanisms. This paradigm shift mirrors the division of labor in human organizations and complex systems, moving from a single, overloaded generalist to a team of orchestrated specialists. The core advantages of this approach are manifold: it allows for modular design and testing, confines knowledge domains to reduce interference, enables concurrent task execution, and provides a structured framework for managing complex, multi-step workflows.

To understand the landscape of these architectures, it is useful to examine a taxonomy of common multi-agent topologies that have crystallized in both research and practice. Each pattern embodies a different philosophy for distributing agency and managing coordination. Figure \ref{fig:taxonomy} shows the taxonomy.

\textbf{Orchestrator--Worker (Manager--Specialist).} This is perhaps the most prevalent and intuitively understandable pattern. A central orchestrator (or manager) agent is responsible for high-level task comprehension, planning, and decomposition. It then delegates discrete sub-tasks to a team of specialized worker agents, such as a researcher, a coder, a critic, or an action executor. The orchestrator synthesizes the workers' outputs to produce a final result. This topology directly mirrors traditional organizational workflows and offers significant benefits for modular evaluation and system robustness, as each specialist can be developed, optimized, and validated independently against its specific role. However, the orchestrator becomes a critical single point of failure and a potential bottleneck for system reasoning.

\textbf{Router--Solver.} This topology emphasizes efficiency and context purity. A router agent acts as a classifier, analyzing an incoming task to determine its type—for example, "code generation," "data analysis," or "content summarization." Based on this classification, the router directs the task to a pre-configured, specialized solver agent best suited for that category. This design drastically reduces context pollution by preventing the mixing of disparate tool sets and knowledge bases within a single agent. It can also improve latency and reduce costs by invoking only the necessary tools for a given task path. The primary challenge lies in designing a robust router and ensuring comprehensive coverage of task types by the solver ensemble.

\textbf{Hierarchical Command Structures.} For managing large-scale, complex projects, a flat orchestrator--worker model may be insufficient. Hierarchical structures introduce multiple layers of management, where top-level managers coordinate mid-level managers, who in turn oversee teams of specialized workers. This recursion allows for the decomposition of exceptionally complex problems into manageable sub-problems distributed across a tree of agents. While this can improve scalability and provide clear lines of responsibility, it also increases coordination overhead, communication latency, and the complexity of failure modes, as errors can propagate through the management chain.

\textbf{Swarm or Market-Like Architectures.} Moving away from rigid, pre-defined hierarchies, swarm architectures employ more decentralized and dynamic coordination. In such systems, a pool of agents, potentially with overlapping or evolving capabilities, can listen to a task broadcast. Agents may then bid for the task based on their self-assessed capability and current workload, or dynamically assume control when their confidence in handling a sub-problem is high. This model can yield emergent specialization and robust load balancing, adapting organically to changing task distributions. However, it introduces significant challenges in governance, predictability, and ensuring consistent quality, as the agent interactions are less prescriptive and more emergent.

The progression from monolithic agents to these orchestrated teams represents a fundamental maturation in AI software architecture. It replaces the paradigm of a solitary, omni-capable intelligence with a socio-technical model of collaborative problem-solving. The choice of topology is not merely an implementation detail but a core architectural decision that trades off between control and emergence, efficiency and robustness, simplicity and scalability. As these systems evolve, the coordination mechanisms themselves—the protocols for communication, negotiation, and joint planning—become the critical substrate for advanced agentic intelligence, marking the transition from simple prompt–response chains to truly goal-directed, organizational-scale systems. Table \ref{tab:failure-modes} shows multi-agent topology failure modes, root causes, mitigation patterns, and detection signals.

\begin{table}[htbp]
\centering
\caption{Multi-agent topology failure modes, root causes, mitigation patterns, and detection signals.}
\label{tab:failure-modes}
\begin{tabular}{>{\RaggedRight}p{2.6cm} >{\RaggedRight}p{2.0cm} >{\RaggedRight}p{2.4cm} >{\RaggedRight}p{3.0cm} >{\RaggedRight}p{3.0cm}}
\toprule
\textbf{Topology} & \textbf{Failure Mode} & \textbf{Root Cause} & \textbf{Mitigation Pattern} & \textbf{Detection Signal} \\
\midrule

Orchestrator–Worker & Silent worker failure & Workers fail without status reporting; orchestrator assumes progress & Heartbeats + explicit ACK/NACK & Missing heartbeats; task timeout without error logs \\
\addlinespace

Orchestrator–Worker & Capability mismatch & Tasks exceed worker's tool/function scope & Semantic capability registry + runtime validation & Repeated tool errors; worker rejection spike \\
\midrule

Router–Solver & Misrouting & Classifier lacks discriminative features for solver specialization & Ensemble routing w/ confidence thresholds + HITL fallback & High solver rejection; repeated re-routing cycles \\
\addlinespace

Router–Solver & Solver overload cascade & Traffic concentration on top solver causes saturation & Load-aware routing w/ backpressure; capacity quotas & Queue depth alerts; path-specific latency degradation \\
\midrule

Hierarchical Command Structures & Command distortion & Goal semantics degrade across layers ('telephone effect') & Signed intent propagation; periodic root-goal alignment & Divergence: leaf actions vs. root objectives \\
\addlinespace

Hierarchical Command Structures & Delegation deadlock & Circular subtask dependencies between peer agents & DAG enforcement; timeout-based revocation & Planning loop timeout; repeated handoffs without progress \\
\midrule

Swarm or Market-Like & Herding behavior & Convergence on local optima suppresses exploration/diversity & Entropy-preserving incentives; anti-correlation penalties & Gini coefficient breach; solution space collapse \\
\addlinespace

Swarm or Market-Like & Strategic manipulation & Sybil collusion distorts market/swarm signals & Bonding curves w/ stake slashing; verifiable delay functions & Abnormal bid retractions; reputation inflation sans contribution \\
\bottomrule
\end{tabular}
\end{table}

\begin{table*}[t]
\centering
\caption{Enterprise Hardening Checklist for Agentic AI Systems}
\label{tab:enterprise_hardening_checklist}
\begin{tabular}{p{2.8cm} p{2.2cm} p{4.2cm} p{3.2cm} p{3.2cm}}
\hline
\textbf{Control Area} & \textbf{Requirement} & \textbf{Implementation Pattern} & \textbf{Verification Method} & \textbf{Evidence Artifact} \\
\hline

Identity \& Access & MUST & Strong identities, short-lived credentials, RBAC, least privilege & Unit tests, IAM audit, red-team & Access/token logs, principal IDs in traces \\

Policy Enforcement & MUST & Central policy gate, policy-as-code, high-risk approvals & Trace queries, policy regression tests & Policy decisions, approval records \\

Tooling \& Integrations & MUST & Typed/versioned interfaces, schema validation, idempotency & Contract/schema tests & Schema registry, tool versions in traces \\

Memory Management & SHOULD & Tiered memory, PII filtering, retention policies & Data governance audit, red-team & Memory access logs, retention records \\

Observability \& Tracing & MUST & E2E structured tracing, standardized metadata & Trace queries, dashboards & Traces with model/prompt/tool versions \\

Budgeted Autonomy & MUST & Caps (tokens/time/cost/tool calls), circuit breakers, safe termination & Stress/chaos testing & Budget metrics, termination logs \\

Data Governance & MUST & Classification, encryption (in transit/at rest), lineage tracking & Compliance audit, regression tests & Lineage records, encryption config \\

CI/CD \& Evaluation & SHOULD & Continuous eval pipeline, regression/safety benchmarks & Automated regression/benchmark tests & Evaluation reports, benchmark history \\

Security Testing & SHOULD & Prompt injection tests, adversarial red-teaming, sandboxing & Red-teaming, penetration tests & Security test reports, incident tickets \\

Change Management & SHOULD & Signed prompts/policies, approval workflows & Config diff checks, audit reviews & Signed hashes, change logs \\

\hline
\end{tabular}
\end{table*}

\section{Industry Platforms and Emerging Architectural Stacks}
The theoretical evolution towards orchestrated multi-agent systems is concretely reflected in the roadmaps and public descriptions of emerging industry platforms. These commercial and open-source ecosystems are actively defining the practical "stacks" for agentic AI, moving beyond isolated frameworks to integrated environments that address the full lifecycle of development, deployment, and governance. For instance, Kore.ai explicitly foregrounds \emph{multi-agent orchestration} as a core capability, emphasizing agent collaboration, supervisor agents, memory, and formal inter-agent protocols \cite{koreai_platform_2026}. Similarly, ZenML references production-ready workflows incorporating ``LangGraph swarms,'' situating agentic patterns within established MLOps pipelines \cite{zenml_platform_2025}. This maturation of orchestration abstractions is further evidenced by extensive LangChain blog coverage on LangGraph and multi-agent application patterns, signaling a shift from prototype to production-grade design \cite{langchain_blog_2026}.

The operationalization of these architectures necessitates rigorous communication protocols and shared context mechanisms. Multi-agent systems require explicit communication contracts, raising core design questions: should message schemas be free-form natural language or structured fields (e.g., task, evidence, constraints)? Is shared memory best implemented as a global scratchpad or as per-agent memory with selective sharing? Furthermore, authority and escalation models must be defined—specifying which agents can commit external actions and when human approval is required. In practice, enterprise systems gravitate towards structured, auditable inter-agent protocols that can be statically validated. This need for contract enforcement is driving the adoption of gateway-oriented architectures, where tool registries, Model Context Protocol (MCP) servers, and centralized policy layers act as the enforcement mechanism for these interactions \cite{truefoundry_agentic_2025}.

However, multi-agent architectures introduce novel and complex failure surfaces that platforms must mitigate. These include \emph{coordination failures} such as deadlocks, redundant work, or directly conflicting actions; \emph{error amplification}, where one agent's hallucination becomes an unchallenged premise for another; \emph{cascading tool calls} that create unbounded action graphs leading to cost overruns or operational incidents; and \emph{policy bypass through delegation}, wherein a restricted action proposed by one agent is executed by a collaborating agent with higher privileges. Effective platforms therefore incorporate architectural mitigations: centralized policy enforcement for all tool calls, strict privilege separation across agent roles, bounded execution budgets, and comprehensive step-level tracing and immutable audit logs \cite{koreai_platform_2026,truefoundry_agentic_2025,langchain_blog_2026}.

A comparative view of selected platforms, summarized in Table~\ref{tab:platforms}, reveals how different ecosystems are shaping the agentic architectural stack through their unique emphases. Salesforce's Agentforce \cite{salesforce_agentforce_2026} frames the agent as the primary interface to enterprise CRM workflows, prioritizing deep integration with business objects and processes alongside robust governance. Kore.ai \cite{koreai_platform_2026} structures its platform around a clear separation between multi-agent orchestration, a unified data connectivity layer, and administrative governance. TrueFoundry \cite{truefoundry_agentic_2025} advocates a "gateway-first" architecture, treating inter-agent and agent-tool traffic as regulated API traffic to enforce sovereignty, isolation, and auditability. ZenML \cite{zenml_platform_2025} positions agents as reproducible, versioned workflows within a unified Directed Acyclic Graph (DAG) orchestration system, extending MLOps discipline to agentic systems. LangChain and LangSmith \cite{langchain_blog_2026} focus on the developer experience, reinforcing deep observability and trace-driven debugging as central to agent engineering.

Two overarching architectural themes crystallize from this landscape. First, \textbf{cross-cutting governance is no longer optional}. Across enterprise platforms, role-based access control (RBAC), immutable audit logs, and policy enforcement are first-class concerns, indicating that agentic systems are being integrated directly into regulated business workflows rather than confined to experimental sandboxes \cite{koreai_platform_2026,truefoundry_agentic_2025,salesforce_agentforce_2026}. Second, \textbf{observability and reproducibility are converging} as core requirements. The emphasis on comprehensive tracing (LangChain/LangSmith, TrueFoundry) and workflow lineage with artifact versioning (ZenML) addresses a shared need: to explain system behavior, reproduce outcomes, and enable safe, iterative improvement \cite{langchain_blog_2026,truefoundry_agentic_2025,zenml_platform_2025}. Together, these themes underscore that the evolution of agentic architecture is as much about instituting software engineering and operational rigor as it is about advancing cognitive capabilities.

\begin{figure}[hbt!]
\centering
\includegraphics[width=0.92\textwidth]{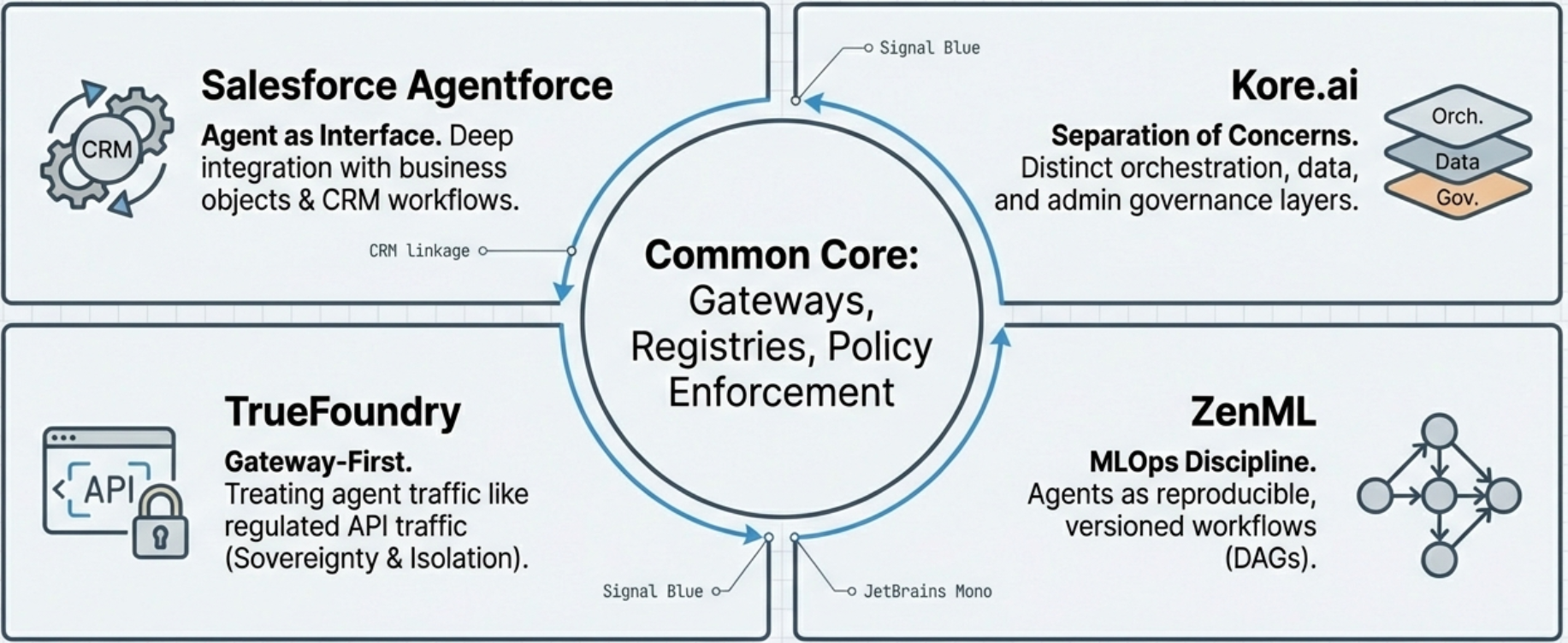}
\caption{The industry is converging on a standard stack: Orchestration + Governance.}
\label{fig:strategy}
\end{figure}

Figure~\ref{fig:strategy} illustrates the industry’s convergence toward a standardized AI platform stack built around orchestration and governance as core layers. This convergence is driven by several key platform strategies, including the use of agents as the primary interaction interface, a clear separation of concerns across system layers, a gateway-first architectural approach, and the adoption of a disciplined MLOps practice to ensure scalability, reliability, and lifecycle governance.

\begin{table*}[t]
\centering
\small
\begin{tabularx}{\linewidth}{@{}p{28mm}p{38mm}p{40mm}X@{}}
\toprule
\textbf{Platform} & \textbf{Primary framing} & \textbf{Notable architectural emphases} & \textbf{Implications for agent architecture} \\
\midrule
Salesforce (Agentforce) \cite{salesforce_agentforce_2026}
& Agentic enterprise CRM platform
& Unifies apps, data, agents; trust \& governance; industry solutions; ROI emphasis
& Agent as interface to enterprise workflows; governance must be first-class; deep integration with business objects and processes \\

Kore.ai (Agent Platform) \cite{koreai_platform_2026}
& Enterprise agent platform for work/service/process
& Multi-agent orchestration; search \& data layer with connectors; no-code + pro-code; observability; security \& governance (RBAC, audit logs, guardrails)
& Strong separation of orchestration, data connectivity, and governance; favors standardized registries and administration \\

TrueFoundry \cite{truefoundry_agentic_2025}
& Agentic deployment + gateways
& AI Gateway; MCP servers; registries; prompt lifecycle; tracing; OpenTelemetry; on-prem/VPC deployment; RBAC and immutable audit logging
& ``Gateway-first'' architecture that treats agent traffic like regulated API traffic; emphasizes sovereignty, isolation, and auditability \\

ZenML \cite{zenml_platform_2025}
& Workflow backbone from pipelines to agents
& Unified DAG orchestration (ML + agents); artifact/environment versioning; caching; infrastructure abstraction; governance and secrets management; SOC2/ISO compliance
& Positions agents as reproducible workflows; pushes agentic systems toward MLOps discipline and lineage-driven debugging \\

LangChain / LangSmith \cite{langchain_blog_2026}
& Agent engineering + debugging
& Deep agents; context management; agent builder templates; traces-to-insights; debugging in production
& Reinforces observability as central; agent development becomes iterative engineering with trace-driven improvement \\
\bottomrule
\end{tabularx}
\caption{High-level comparison of publicly described emphases in selected agentic AI platforms and ecosystems. Entries summarize what the referenced materials emphasize rather than benchmarking performance.}
\label{tab:platforms}
\end{table*}

\section{Open Challenges and Research Directions}
The transition from prompt–response interfaces to goal-directed, agentic systems represents a fundamental shift in software architecture, introducing profound challenges that extend beyond mere tooling. While current platforms provide essential scaffolding for orchestration, tool integration, and memory, several critical areas remain unresolved. These challenges are not merely incremental improvements but may necessitate new abstractions and foundational research to ensure the robust, safe, and scalable deployment of autonomous agents in real-world settings.

A primary obstacle lies in achieving verifiability and formal guarantees over agent behavior. The inherent stochasticity and opaque reasoning processes of large language models (LLMs) make it difficult to verify the correctness, safety, or compliance of an agent's decisions post-hoc. Promising research directions aim to impose structure on this uncertainty. These include developing typed action plans that constrain the agent to produce executable sequences adhering to a predefined schema, enabling automated validation of plan consistency and safety preconditions. The concept of proof-carrying actions further requires agents to attach actionable evidence—such as retrieval citations, tool outputs, or confidence scores—to their key decisions, creating an audit trail. Complementing these, conformance testing frameworks, including regression suites across complex tool graphs and multi-agent workflows, are essential for maintaining reliability as systems evolve. In enterprise contexts, these technical approaches must seamlessly connect to established audit, compliance, and governance mechanisms, a requirement underscored by commercial platform designs \cite{koreai_platform_2026, truefoundry_agentic_2025}.

Closely related is the challenge of interoperability and compositional safety. As agentic ecosystems grow, a recurring architectural need is for standardized protocols for agent-to-tool and agent-to-agent communication, complete with schema validation and policy controls. Industry movements, such as the adoption of the Model Context Protocol (MCP) for tool discovery and the development of inter-agent communication channels, signal a push toward shared standards \cite{truefoundry_agentic_2025, koreai_platform_2026}. This raises a fundamental research question: what constitutes the minimal, stable contract—encompassing capabilities, permissions, and error semantics—that enables safe and efficient composition of heterogeneous agents and tools across organizational boundaries? A vision akin to "REST for agents" is emerging, but one that must address challenges of trust, dynamic discovery, and negotiation that surpass those of traditional web services.

The increased autonomy granted to agents necessitates rigorous frameworks for safety and graceful degradation. Agentic systems must be designed to fail safely, avoiding catastrophic failures when faced with novelty, ambiguity, or error. Key architectural patterns include establishing clear escalation paths to route uncertain cases to human oversight or fallback deterministic workflows. Implementing budgeted autonomy—enforcing strict quotas on tokens, reasoning time, tool calls, or cost—is crucial to bound operational risk and financial exposure. Furthermore, a sandbox-first execution paradigm, where agents can simulate or evaluate the potential side effects of actions in a controlled environment before commitment, provides a critical safety layer. These principles align with the enterprise focus on governed collaboration, as seen in platforms that emphasize human-in-the-loop controls and administrative governance rather than unconstrained automation \cite{salesforce_agentforce_2026, koreai_platform_2026}.

Expanded capabilities for persistent memory and broad tool access intensify concerns over data governance, privacy, and sovereignty. The architectural patterns for agentic systems must embed privacy-by-design principles from the ground up. This includes strategies to minimize stored personal data, favoring the retention of hashed references, anonymized summaries, or ephemeral context over raw sensitive information. Policy-aware retrieval mechanisms must ensure that an agent's memory and tool access are dynamically filtered based on the current user's permissions and the specific purpose of the task. For regulated sectors, architectural support for on-premises or virtual private cloud (VPC) deployment is non-negotiable, ensuring that data never leaves a customer's controlled domain. This sovereignty-by-design approach is explicitly foregrounded by platforms targeting regulated industries and is reinforced as a key leadership concern in enterprise adoption \cite{truefoundry_agentic_2025, zenml_platform_2025, bain_home_2026}.

Finally, the evolution of agentic architecture will inevitably intersect with the physical world through embodied and cyber-physical agents. While this paper focuses on software agents, the architectural principles must extend to robotics and systems where agents actuate change in real environments. Research in adjacent domains, such as using LLMs for simulating virtual patients or robotic task planning, highlights this convergence \cite{cmu_ri_home_2026}. In cyber-physical settings, the architectural paradigm must hybridize: low-level, reactive safety layers implementing hard constraints (drawing from classical control theory and hybrid systems) must coexist with, and override, high-level deliberative planners \cite{brooks_1986, russell_norvig_2020}. This returns the field to classical questions in AI architecture but with new, neural-centric reasoning components, demanding novel frameworks for verification, real-time performance, and safety assurance that bridge the discrete symbolic and continuous physical realms.

\section{Threats to Validity}
As an analytical survey focused on the architectural evolution of agentic AI, this paper is subject to several methodological and contextual limitations that must be acknowledged to properly bound its conclusions and guide future research.

A primary concern is the potential for grey literature bias. The analysis heavily incorporates technical blogs, white papers, and platform documentation from leading vendors and consultancies. By nature, such sources are prone to emphasizing capabilities, strengths, and strategic visions while omitting limitations, failures, and implementation hurdles. To mitigate this bias, our methodology restricts its claims to explicit, descriptive statements found within these materials regarding system design, component interfaces, and stated priorities. We consciously use these sources not as evidence of efficacy or performance, but as artifacts that reveal industry-wide architectural trends, consensus patterns, and emergent design priorities. This approach allows us to map the conceptual landscape as presented by key ecosystem builders, while acknowledging that a complete picture requires complementary data from peer-reviewed research and independent empirical studies.

The field is also characterized by extremely rapid evolution in both tooling and terminology. New frameworks, agent "runtime" environments, and proprietary platforms are announced frequently, often with shifting feature sets and overlapping definitions of core concepts like "memory," "tools," or "orchestration." While this dynamism poses a challenge for any contemporary analysis, we assert that the underlying architectural primitives discussed in this paper—such as the control loop for reasoning and acting, mechanisms for stateful memory, interfaces for tool abstraction, and layers for governance—exhibit relative stability. These components represent fundamental computational requirements for building goal-directed systems, irrespective of their specific instantiation in any given library or product. Consequently, while specific APIs and commercial offerings will inevitably change, the conceptual framework of agentic architecture presented here is intended to provide a durable lens for understanding future developments.

Finally, the scope of this paper is deliberately constrained. It focuses principally on the software architecture for digital agents operating within information systems, explicitly excluding the complex domain of embodied robotics and cyber-physical systems, save for a brief discussion of future intersections. Furthermore, this work is a survey and analytical synthesis, not an empirical evaluation. It does not present performance benchmarks, comparative efficiency metrics, or formal verification results for the architectures described. This scope limitation means the paper identifies what is being built and how it is conceptually structured, but does not establish how well different patterns perform under specific conditions. Future work is essential to complement this architectural map with rigorous, controlled experiments measuring robustness, scalability, and safety, as well as longitudinal analyses of real-world deployment incidents. Such empirical research will be crucial for transforming the current landscape of promising architectural patterns into a mature engineering discipline with known trade-offs and validated best practices.

\section{Conclusion}
This analysis of the agentic AI landscape reveals that the transition from simple prompt–response interfaces to autonomous systems represents a fundamental reorganization of software architecture, one that transcends incremental improvements in prompt engineering. At its core, this evolution marks the emergence of a new architectural paradigm: the construction of goal-directed control loops where large language models function as cognitive kernels for high-level planning and reasoning. These kernels are then systematically embedded within a supporting constellation of specialized components—persistent and episodic memory systems, tool abstraction layers, policy enforcement engines, and comprehensive observability frameworks—that grant them the capacity to act reliably and accountably over time.

This modern instantiation is deeply rooted in the enduring abstractions of classical agent theory, which emphasized the primacy of state, intent, deliberation, and reactivity. Contemporary architectures explicitly instantiate these concepts, translating them into components for state management (via vector databases and session caches), intent decomposition (through planner modules), and reactive supervision (via guardrail models and circuit breakers). The significant advancement lies in how modern enterprise practice is hardening these theoretical constructs. Governance-by-design is now a non-negotiable requirement, manifesting in architectural features like fine-grained Role-Based Access Control (RBAC) for tools and memory, immutable audit logs for full traceability, and native policy hooks for compliance validation. Simultaneously, the imperative for workflow reproducibility is driving the integration of software engineering best practices, including versioning for prompts, tool specifications, and agent configurations, as well as the use of containerized environments to ensure consistent execution.

A review of contemporary platforms and frameworks indicates a notable convergence in architectural blueprints. While implementations vary, a common pattern has crystallized around core infrastructural services: centralized registries for agents, tools, and skills; API gateways managing authentication, routing, and rate-limiting; standardized schemas (e.g., OpenAPI, MCP) for tool interoperability; and sophisticated orchestration engines capable of managing complex, hierarchical multi-agent workflows—all operating under a unified umbrella of auditable controls \cite{koreai_platform_2026, truefoundry_agentic_2025, zenml_platform_2025, langchain_blog_2026, salesforce_agentforce_2026}. This convergence signals the maturation of agentic AI from a research prototype into a bona fide category of enterprise software, demanding and developing its own dedicated stack.

Looking ahead, the next phase of evolution will be defined by the challenge of safe composability. As agentic systems proliferate and must interact across organizational boundaries—much like microservices in a distributed system—the need for interoperable protocols and stronger verification techniques becomes paramount. The future trajectory points toward the development of a "service mesh for agents," featuring standard contracts for capability discovery, trust negotiation, and verifiable credential exchange. Furthermore, ensuring reliability at scale will require advances in formal and probabilistic verification methods to provide guarantees on agent behavior, moving beyond monitoring and toward provable constraints. The historical parallel is clear: just as the web services revolution was enabled by shared interface standards (SOAP, REST) and layered governance (API management), the widespread, trustworthy deployment of autonomous agentic systems hinges on the community's ability to establish equivalent foundations for safe, scalable, and intelligible composition. The evolution from prompt–response to goal-directed systems is, therefore, not merely a technical shift but the genesis of a new software engineering sub-discipline, one that must seamlessly blend insights from artificial intelligence, distributed systems, and safety-critical design.

\end{document}